\documentclass[%
 reprint,
 amsmath,amssymb,
 prl,
]{revtex4-2}

\usepackage{float}

\usepackage{graphicx}
\usepackage{dcolumn}
\usepackage{bm}
\usepackage{amsthm,amsmath,amssymb,tikz}
\usetikzlibrary{shapes.misc,shapes,decorations,circuits.logic.US,circuits.logic.IEC,fit,external,shapes.gates.logic.US,trees,positioning,arrows,calc,shapes.geometric,shapes.symbols}
\tikzstyle{loosely dashed}=[dash pattern=on 4pt off 8pt]
\tikzstyle{loosely dashed2}=[dash pattern=on 6pt off 8pt]

\tikzset{
    connection/.style={
        draw = none,
        circle,
        radius=5pt,
    },
    uptri/.style={
        isosceles triangle,
        isosceles triangle apex angle=60,
        shape border rotate=90
    },
    downtri/.style={
        isosceles triangle,
        isosceles triangle apex angle=60,
        shape border rotate=-90
    },
}

\begin{document}


\title{Transformations of mixed spin-class Ising systems}

\author{Joost Kruis}
\affiliation{%
 University of Amsterdam, Psychological Methods Department
}%




\date{\today}

\begin{abstract}
In the many fields in which the Ising model is applied nowadays, the spin variables are often assumed to be of spin-class $\{-1,1\}$ or $\{0,1\}$, even though for any mix of binary real valued spin-classes a proper Ising model distribution exists. Here we show that in their basis all these spin-classes are the same, as a simple expressions exist that allow us to transform the variables from one particular mix of spin-classes to any other combination, without changing the probabilities for the system states as a whole.

\end{abstract}

\maketitle


\section{}

\section{}

\vspace{-50pt}

The Ising model\cite{1920lenz,1925ising} is one of the most popular ways to model the structure of associations between a set of binary variables. Originally from statistical physics, it was formulated to describe the ferromagnetic behaviour of a system of atomic spins. In this application a spin variable denoted by $\sigma \in \{-1,+1\}$ indicates the direction of the spin, and takes value $\sigma = -1$ if the atom spins in a downward direction, and $\sigma = +1$ if the atom spins in an upward direction. Such that for a collection of spins $(\boldsymbol{\sigma})$ that make up a system, these states are distributed as:

\begin{align}
    \begin{split}
        p(\boldsymbol{\sigma}) &= \frac{1}{Z} \exp\left(\beta\left[\sum\limits_{\langle i, j \rangle} a_{ij} \sigma_i \sigma_j + \mu \sum\limits_i b_i \sigma_i \right]\right) \\
        \\
        Z &= \sum\limits_{\boldsymbol{\sigma}} \exp\left(\beta\left[\sum\limits_{\langle i, j \rangle} a_{ij} \sigma_i \sigma_j + \mu \sum\limits_i b_i \sigma_i \right]\right) \, ,
    \end{split}
    \label{eq:01}
\end{align}

\noindent where $\sum_{\langle i, j \rangle}$ denotes the sum over all distinct pairs of $i$ and $j$, $a_{ij} \in \mathbb{R}$ denotes the strength of the pairwise interaction between spin $i$ and $j$, $b_{i} \in \mathbb{R}$ denotes the strength of the external magnetic field on spin $i$, $\beta \in \mathbb{R}_{\geq 0}$ is the inverse thermodynamic temperature of the system, and $\mu \in \mathbb{R}$ the magnetic moment scaling the influence of the external field as a function of its strength and orientation with respect to the system.

While its intended application was later rejected due to incompatibility with quantum mechanics it has remained highly popular and one of the most studied models in modern statistical physics\cite{2005niss}. As it is capable of capturing complex phenomena, by modelling the joint distribution of binary variables as a function of main effects and pairwise interactions \cite{2018marsman_borsboom}, it has also attracted attention outside physics\cite{1999stutz_williams} and has been used in fields such as genetics\cite{2015fierst_phillips}, educational measurement\cite{2015marsman_maris}, psychology\cite{2018epskamp_maris}, and decision-making\cite{2014verdonck_tuerlinckx}. In the statistics literature the distribution without $\beta$ and $\mu$ is known as the \textit{quadratic exponential binary distribution}\cite{1972cox, 1994cox_wermuth}.

Perhaps the most recent field to embrace the Ising model has been psychology, where is has been positioned as an alternative explanation for the then dominant, but theoretically problematic, latent variable perspective. By establishing the mathematical equivalence between the Ising model and an often used class of binary latent variable models\cite{2006van-der-maas_dolan,2018epskamp_maris, 2015marsman_maris, 2018marsman_borsboom, 2016kruis_maris}, the historical success of fitting these latent variable models to data could be accounted for, while providing a theoretically more plausible explanation of the observed associations. However, instead of binary variables $\sigma \in \{-1,+1\}$, common practice has been to use variables $\sigma \in \{0,1\}$, indicating for example that an item was answered either incorrect $(\sigma = 0)$ or correct $(\sigma = 1)$, or to indicate that a symptom is absent $(\sigma = 0)$ or present $(\sigma = 1)$. 

This is no problem as we obtain a proper distribution through the normalising constant $Z$, also known as the partition function, that sums over all possible configurations of the system. As such, when using the same set of values for $\mathbf{A}$ and $\mathbf{b}$ the probability distribution for all $\boldsymbol{\sigma} = 1$ will be different depending on whether we use $\sigma \in \{-1,1\}$ or $\sigma \in \{0,1\}$. There exists however a linear transformation of the Ising model with $\sigma^x \in \{-1,1\}$ to that of $\sigma^y \in \{0,1\}$ (we add the superscript for clarity). Recognising that $\sigma^x = 2 \, \sigma^y - 1$ and substituting in equation \ref{eq:01} we can derive that by letting $a_{ij}^y = 4a_{ij}^x$ and $b_{i}^y = 2b_{i}^x - 2a_{i+}^x$, where $a_{i+}$ denotes the sum over all pairwise interactions from spin $i$, we obtain that $\boldsymbol{\sigma}^x$ and $\boldsymbol{\sigma}^y$ have the same probability distribution over their states. 

Although traditionally applications of the Ising model have assumed that all spins in a single system can only take the same two values, nothing in the expression of the model precludes the system to contain both $\sigma \in \{0,1\}$ and $\sigma \in \{-1,1\}$ variables, or any collection of $\sigma_i \in \{\sigma_{i1} \in \mathbb{R},\sigma_{i2} \in \mathbb{R}\}$ spins for that matter, for as long as $Z$ sums over all possible states of the system we would obtain a proper distribution. In other words, if we define a spin-class as the set of two real-valued numbers a spin variable can take, we can envision an Ising model in which each node has a different spin-class that has a valid probability distribution as long as we sum over all possible configurations of these spins in the normalising constant. In itself this might be a trivial and even bothersome statement, given that all this does is introduce an endless amount of new models, that at first sight might even complicate things particularly for fields outside of physics. However, we believe that with our second statement, and the goal of this Letter, we can actually make show why mixed spin-class Ising models are a nice addition to our repertoire. 

In this Letter we report the expressions that allow us to transform an Ising model from any collection of real valued binary spin-class variables to any other combination of spin-classes, and discuss the implications of this capability. These expressions entail first transforming all spins from their original spin-class to a $\sigma \in \{0,1\}$ class, from which we then can transform them to their target class. This shows that estimation of the Ising model, particularly outside of physics, can still be done with all the current methods. Furthermore, trough transformations of the spins to different spin-classes one can for example investigate under which spin-class configuration the external field of a system is minimal, or impose more theoretically inspired constraints on the model and evaluate there influence on for example centrality measures. 
\vspace{-15pt}
\subsubsection{Transforming spin-classes} 
\vspace{-10pt}
The first key insight that allows us to transform mixed binary spin-class systems at will, while keeping the probability distribution over the states the same, is that we can transform any set of two real valued numbers $\sigma_i = \{\sigma_{i1}, \sigma_{i2}\}$ to a set $\sigma_i^{\ast} = \{0,1\}$ by the operation $\sigma_i^{\ast} = \frac{\sigma_i-\sigma_{i1}}{\sigma_{i2}-\sigma_{i1}}$. As the first operation sets the first element of $\sigma_i, \sigma_{i1}$ to zero, and the second element $\sigma_{i2}$ to $\sigma_{i2} - \sigma_{i1}$, such that the second operation divides $\sigma_i = \{0, \sigma_{i2} - \sigma_{i1}\}$ by $\sigma_{i2} - \sigma_{i1}$, so unless $\sigma_{i1} = \sigma_{i2}$ the first element of $\sigma_i$ will remain zero, and the second element will become one. Note that in the event that $\sigma_{i1} = \sigma_{i2}$, we are not dealing with a binary variable anymore and hence this situation is not applicable to the current setting. The second insight is that this first process is reversible, and hence we can transform the set $\sigma_i^{\ast} = \{0,1\}$ to any set of two real valued numbers $\sigma_i = \{\sigma_{i1}, \sigma_{i2}\}$ with the operation $\sigma_i = \sigma_i^{\ast} \cdot (\sigma_{i2} - \sigma_{i1}) + \sigma_{i1}$. In which the first operation sets the second element of $\sigma_i^{\ast}$ to $(\sigma_{i2} - \sigma_{i1})$, and the second operation adds $\sigma_{i1}$ to both elements in the set. The next step is to combine these insights. 

If we let $\mathbf{x}$ denote the collection of the original spin-classes of the variables in the system, and $\mathbf{y}$ the collection of target spin-classes for these variables. We can use the following expression to transform any variable in $\mathbf{x}$ to their target spin class in $\mathbf{y}$: 

\begin{align}
    \begin{split}
        x_{ij} \rightarrow y_{ij} &= \frac{x_{ij} - x_{i1}}{x_{i2} - x_{i1}} \cdot \left(y_{i2} - y_{i1}\right) + y_{i1} \, ,
    \end{split}
    \label{eq:02}
\end{align}

\noindent where $i$ denotes the specific variable in the system, and $j$ the index of the value. For example, if spin $i \in \{2,4\}$, $x_{i1} = 2$ and $x_{i2} = 4$. A closer inspection of equation \ref{eq:02} shows that the first part transforms the $x_{i1}$ to $0$, and $x_{i2}$ to $1$, and the second part transforms them to the target spin-class. To obtain the required transformations of $\mathbf{A}$ and $\mathbf{b}$, we can plug equation \ref{eq:02} into equation \ref{eq:01}, and derive the expressions for these transformations. 

As a first step we will transform $\mathbf{A}$ and $\mathbf{b}$ to the situation in which all variables are in, what we term, the standard spin-class $\mathbf{s}$, where $s_{i1} = 0$ and $s_{i2} = 1$. Realising that $\beta$ and $\mu$ are constants we can simplify the derivation by setting $\mathbf{A}^x = \beta \cdot \mathbf{A}$ and $\mathbf{b}^x = \beta \cdot \mu \cdot \mathbf{b}$, where the superscript $x$ denotes that these are the parameters in the original class. Let $\mathbf{A}^s$ and $\mathbf{b}^s$ contain the parameters when the variables are transformed to the standard class, which we obtain using the following expression:  

\begin{align}
    \begin{split}
        a_{ij}^s &= a_{ij}^x \, \left(x_{i2}-x_{i1}\right) \, \left(x_{j2}-x_{j1}\right) \\
        \\
        b_i^s &= b_i^x \, \left(x_{i2}-x_{i1}\right) + \sum_j a_{ij}^x \, \left(x_{i2}-x_{i1}\right)  \, x_{j1}     
    \end{split}
    \label{eq:03}
\end{align}

One can use this equation to quickly verify the parameter transformation described in the first part of this Letter. The second step is to transform all variables from the standard class $\mathbf{s}$ to their target class $\mathbf{y}$. Let $\mathbf{A}^y$ and $\mathbf{b}^y$ contain the parameters when the variables are transformed from the standard class to their target class, which we obtain using the next expression: 

\begin{align}
    \begin{split}
        a_{ij}^y &= \frac{a_{ij}^s} {\left(y_{i2}-y_{i1}\right) \, \left(y_{j2}-y_{j1}\right)} \\
        \\
        b_i^y &= \frac{b_i^s}{\left(y_{i2}-y_{i1}\right)} + \sum_j \frac{-y_{j1} \, a_{ij}^s}{\left(y_{i2}-y_{i1}\right) \, \left(y_{j2}-y_{j1}\right)}
    \end{split}
    \label{eq:04}
\end{align}

Note that if one would apply this expression for to the example in the first part, as $\mathbf{y} = \mathbf{s}$ and hence, $y_{i1} = 0$ and $y_{i2}-y_{i1} = 1$, the expression \ref{eq:04} will reduce to $a_{ij}^y = a_{ij}^s$ and $b_{i}^y = b_{i}^s$. We can now also take out $\beta$ and $\mu$ again by setting $\mathbf{A} = \frac{\mathbf{A}^y}{\beta}$ and $\mathbf{b} = \frac{\mathbf{b}^y}{\beta \cdot \mu}$. 
\vspace{-15pt}
\subsubsection{Unchanged estimation} 
\vspace{-10pt}
Naturally, one could have written out expressions \ref{eq:03} and \ref{eq:04} as a single expression. We chose not to do this for two reasons, the first being that substituting all $a_{ij}^s$ and $b_i^s$ in equation \ref{eq:04} with the appropriate expression in equation \ref{eq:03} would mean a more cluttered expression. More importantly, it shows a characteristic of the expressions that is particularly interesting for those applications where the model has to be estimated from data. 

Often the Ising model is estimated using a pseudo-likelihood approach where one optimises the conditional distribution of one spin given all other spins using logistic regression, where the sum of the log-likelihoods has been shown to be a reasonable approximation for the full likelihood\cite{1975besag,2014van-borkulo_borsboom}. Applications in which the data is traditionally coded as the standard class $\sigma \in \{0,1\}$. As the expressions for the transformation show that at some point all spins will be in this standard class, it means that application of these transformation does not require us to come up with new estimation procedures. 
\vspace{-15pt}
\subsubsection{Changed interpretation} 
\vspace{-10pt}
As transforming the parameters and spins of an Ising model to a different class using the expressions above does not change anything about the probability distribution over the states, for those who purely study the analytic expression itself not much changes as well. The same goes for applications in which the actual values of the spin-class can be observed for the variables, unless one could not estimate the Ising model in this particular class before, not much will change. However, for those situations in which the Ising model is used as a formal model of some phenomenon one wants to explain as a function of main effects and pairwise interactions there are several ramifications of the possibility for this transformation.  

For one, the $b$ parameter that represents the external magnetic field in the traditional Ising model application is often interpreted as a measure of the amount of variation in the state of a spin cannot be explained by the other spins in the system\cite{2020kruis_maris}, or as a tendency of the spin to be in a particular state as a function of some internal characteristic\cite{2014van-borkulo_borsboom}. As such, inferences from this parameter are for example made about the necessity to look for other variables that are not previously taken into account, or the extent to which can expect to change the state of a variable as a function of interventions. It is clear from equation \ref{eq:03} and \ref{eq:04} that the value of $b$, and hence our inferences, will depend on the spin-class in which we view the variables.

\begin{figure}[H]
    \centering
    \includegraphics[width=0.5\textwidth]{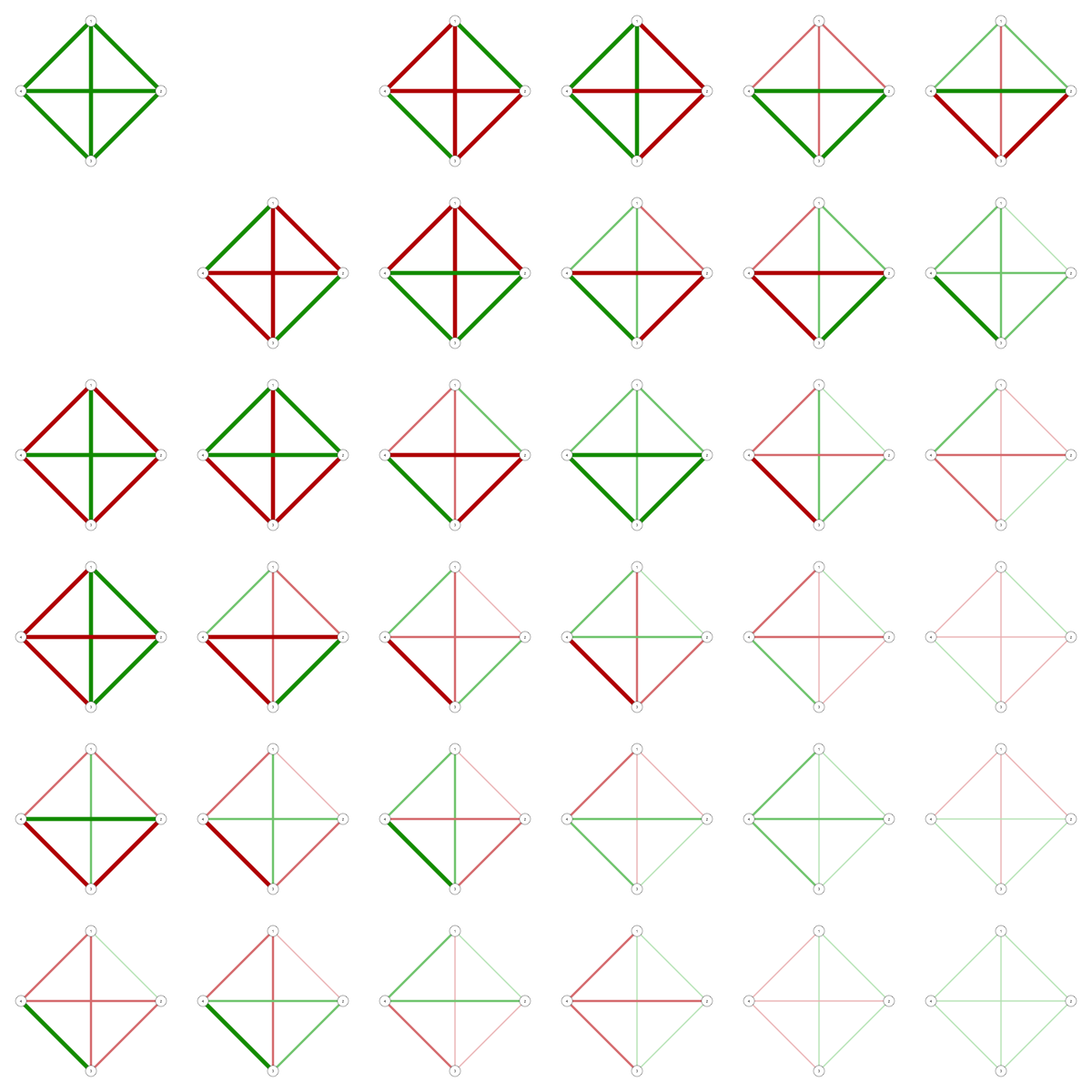}
    \caption{\textbf{A collection of Ising networks with the same probability distributions over the states} \\ Green edges denote relations for which $a_{ij} > 0$, red edges denote relations for which $a_{ij} < 0$, the thickness of the edges denote the relative strength of the relation with respect to the other edges in the same graph, whereas the opacity of the edge denotes the strength of the relation with respect to all graphs in the figure.}
    \label{fig:01}
\end{figure}

Secondly, a large number of applications now-a-days focus heavily on the representation of an Ising model as a graph, and graph characteristics such as centrality measures. As such measures are often a function of the strength of the relationships between two variables as given by $a_{ij}$, once again the calculated centrality measures will be a function of the spin-classes. For example, in Figure \ref{fig:01} we demonstrate this for a simple Ising network, shown in the upper left corner of the figure, with 4 nodes that are all in the standard class $\sigma \in \{0,1\}$, and all $a_{ij} = 1$, and all $b_i = 0$. 

\vspace{-5pt}

Although the other graphs in the figure look completely different, and would lead to different calculations of some centrality measures, they have exactly the same probability distribution over their states as the original network. However, instead of all spins being in the standard class each graph represent a different combination of spins that are either in $\sigma \in \{0,1\}$, $\sigma \in \{-1,1\}$ or $\sigma \in \{-1,0\}$. Clearly, in cases where the visual representation of the model is interpreted, the actual spin-classes in which the variables are, do thus matter for the inferences made from the model.

One should thus be always be cautious when making inferences about a system distributed as an Ising model from visual inspection of the graph, or estimated parameter values, alone. Although purely focusing on the distribution of the system states might be the best solution to tackle unwarranted inferences, we understand that the need for providing a visual aid sometimes requires one to display the model as a graph. In these cases one should use theory to establish the expected spin classes of the variables, for example based on some function of ordinal constraints believed to exist in the literature.
\vspace{-15pt}
\subsubsection{New applications} 
\vspace{-10pt}
Luckily, the possibility of these transformations also provides for some interesting opportunities. For example, in cases where $b$ is interpreted as some unmeasured external force one case try to find the particular configuration of the spin-classes for the system that minimises the influence of $b$. If we assume that we estimated the model for all spins in the standard class, and define $|b^y| = \sum |b_i^y|$, we can try to solve the system of equations in equation \ref{eq:04} for each spin such that it minimises $|b^y|$. In that sense providing a lower boundary on the variance not explained by the pairwise interactions. Likewise, one can also try to find the configuration of spin-classes such that (for groups of spins) all pairwise interactions are as close as possible to being the same magnitude, which might proof useful for clique detection. 
\vspace{-15pt}
\subsubsection{Summary} 
\vspace{-10pt}
The Ising model is making a steady advance into diverse fields of science as a tool to model associations between binary variables as a function of main effects and pairwise interactions. For many current applications of the model, the quantities $\mathbf{A}$ and $\mathbf{b}$ are unobserved and have to be estimated as parameters from data. This is generally done with logistic regression assuming that all spins are in the $\sigma = \{0,1\}$ spin-class, i.e., the set of two values the variable can take. However, as analytically for each combination of spin-classes a equivalent representation exists using a whole other combination of spin-classes, it is important to realise that the particular spin-class assumed will influence the inferences made from the model based on parameters or visual inspection of the model as a graph alone.

\bibliography{MS_BIB}

\end{document}